\documentclass[final,3p,8pt]{elsarticle}

\usepackage{amssymb}
\usepackage{array}
\usepackage{mathrsfs}
\usepackage{amscd}
\usepackage{amsmath}
\usepackage{amsfonts}
\usepackage{amsthm}
\usepackage{psfrag}
\usepackage{textcomp}
\usepackage{bm}
\usepackage{url}
\usepackage{dsfont}
\usepackage{algorithm}
\usepackage{algpseudocode}

 \usepackage{lineno}

\journal{\url{https://arxiv.org/}}

\usepackage{amssymb}

\biboptions{square,sort&compress}

\usepackage[figuresright]{rotating}
\usepackage{moreverb}
\usepackage{amssymb}
\usepackage{array}
\usepackage{mathrsfs}
\usepackage{amscd}
\usepackage{amsmath}
\usepackage{amsfonts}
\usepackage{amsthm}
\usepackage{psfrag}
\usepackage{textcomp}
\usepackage{bm}
\usepackage{color}
\usepackage{float}
\usepackage[applemac]{inputenc}
\usepackage{afterpage}
\usepackage{tabularx}
\usepackage{booktabs}
\usepackage{romannum}
\usepackage{multirow}
\usepackage[normalem]{ulem} 
\usepackage{cancel}

\begin{document}
	

\begin{frontmatter}




\title{Hyperuniformity of quasiperiodic tilings generated by continued fractions}


\author[upv1,upv2]{Mario L\'azaro\corref{cor}}
\author[upv2]{Luis M. Garc\'ia-Raffi}

\cortext[cor]{Corresponding author. Tel +34 963877000 (Ext. 76732), \textnormal{malana@upv.es} }
\address[upv1]{Department of Continuum Mechanics and Theory of Structures\\}
\address[upv2]{Instituto Universitario de Matem\'atica Pura y Aplicada\\
Universitat Polit\`ecnica de Val\`encia (Spain)}


\begin{abstract}
	
Hyperuniformity is a property of certain heteroneous media in which density fluctuations in the long wavelength range decay to zero. In reciprocal space this behavior translates into a decay of Fourier intensities in the range near small wavenumbers. In this paper quasiperiodic tilings constructed by word concatenation are under study. The lattice is generated from a parameter given by its continued fraction so that quasiperiodicity emerges for infinite when irrational generators are into consideration. Numerical simulations show a quite regular quadratic decay of Fourier intensities, regardless of the number considered for the generator parameter, which leads us to formulate the hypothesis that this type of media is strongly hyperuniform of order 3. Theoretical derivations show that the density fluctuations scale in the same proportion as the wavenumbers. Furthermore, it is rigorously proved that the structure factor decays around the origin according to the pattern $S(k) \sim k^4$. This result is validated with several numerical examples with different generating continued fractions.
	
\end{abstract}


\end{frontmatter}


\section{Introduction}


Hyperuniformity is a property of a spatial distribution in which there are fewer density fluctuations at long length scales compared to a random distribution with the same number of points. In other words, hyperuniform systems have a more ordered structure than a typical random system, while still being statistically homogeneous. Quasicrystals (as Crystals) are hyperuniform~\cite{Torquato-2003}. For hyperuniform latticces, the structure factor $S(k)$ is a smooth function that tends to zero as the wavenumber $k$ tends to zero following a power law, according to the equation 
\begin{equation}
	S(k)  \sim k^\gamma
	\label{eq000a}
\end{equation}
A consistent approach to standard cases characterized by smooth $S(k)$ and quasicrystals with dense but discontinuous $S(k)$ in 1D can be achieved by defining $\gamma$ based on the integrated Fourier intensity
\begin{equation}
	Z(k) = 2 \int_0^k S(\kappa) \, d \kappa
	\label{eq000b}
\end{equation}
The integral is multiplied by 2 to be consistent with the definition for higher dimensions, where $\kappa$ is viewed as a radial coordinate. 
For quasiperiodic lattices, $Z(k)$ is monotonically increasing and for $k$ sufficiently small it can be plotted as bounded between to power-law curves verifying that
\begin{equation}
	d_1 \, k^{\gamma+1} < Z(k) < d_{2} \, k^{\gamma + 1}
	\label{eq00d}
\end{equation}
for some constant coefficients $d_1$ and $d_2$ and for some $\gamma$. In such case $\gamma$   is said to be order of hyperuniformity and the cumulative intensity function obeys a power law of order $1 + \gamma$, which is symbolized as
$$
Z(k) \sim k^{\gamma +1} \ , \qquad \text{as} \ k \to 0
$$ 
Another measure of hyperuniformity is given by the local number variance of particles within a window of radius $R$ (an interval of length 2R in the 1D case), denoted by $\sigma^2 (R)$ (order metric context). If its growth is more slowly than the window volume (proportional to $R$ in 1D) in the large-$R$ limit, the system is hyperuniform. For any 1D system the scaling of $\sigma^2 (R)$  for large $R$ is determined by  $\gamma$ as follows: (class I, $\gamma>1$) strongly hyperuniform, (class II, $\gamma =1$) logarithmic hyperuniform and (class III, $0<\gamma<1$) weakly hyperuniform. Finally, for $\gamma<0$ we have the anti-hyperuniform class ~\cite{Torquato-2018a}. \\


In recent years, there has been significant research interest in hyperuniformity in quasiperiodic tilings, which are complex arrangements of tiles with long-range order but no translational symmetry. Studies have shown that certain types of quasiperiodic tilings exhibit hyperuniformity \cite{Torquato-2017,Torquato-2018a}, which has important implications for the physical and mechanical properties of these materials. 
Since hyperuniformity directly invokes particle order in the long wavelength range, the study of tile density depends on the generation pattern of these lattices. Thus, Or\u{g}uz et al. \cite{Torquato-2017} have studied the hyperuniformity order in quasiperiodic lattices generated by projection, showing that it depends on the type of strip used. The so-called ideal strips result in hyperuniformity exponents of $\gamma = 3$. 
Other important case of quasiperiodic 1D structures are those generated by substitution rules (or inflation rules). It turns out that  there is a strong relationship between the scaling in Fourier intensities and density fluctuations in the limit tiling  \cite{Baake-2011,Baake-2012,Godreche-1992}. In particular,  the eigenvalues of the substituion matrix   play an important role in this relationship \cite{Godreche-1993, Luck-1993}. The hyperuniformity of substitution quasiperiodic tilings has been discussed in detail in the reference \cite{Torquato-2018b}, showing that the power-decay law of Fourier intensities strongly depends on the nature of the substitution matrix. Fuchs et. al. \cite{Vidal-2019} have found log-periodic oscillations of the broadening of Landau levels in the presence of a potential with discrete scale invariance, determining exactly the hyperuniformity exponent and the period of such oscillations. \\


In this paper we present a comprehensive analysis of the hyperuniformity of quasi-periodic lattices based on word concatenation and generated by continuous fractions. By making use of the recursive nature of these systems, iterative expressions for both Fourier intensities and density fluctuations can be obtained analytically. Both are a function of the different coefficients that form the continued fraction. Analytical expressions for the decreasing pattern of Fourier intensities with wavenumbers for quasiperiodic tilings generated by the so-called Metallic means and by periodic continued fractions are  derived in detail, showing good agreement with the numerical examples. Furthermore, it is also proved  that the global hyperuniform behavior of the structure factor is $S(k) \sim k^4$.

\section{Quasiperiodic tilings generated by concatenation}

Let us consider two segments (tiles) of lengths $A$ and $B$ and a real number $\alpha \in \mathbb{R}$ defined in the range $0 < \alpha \leq 1$.  Let the sequence $[0;a_1,\ldots,a_n]$ be the continuous fraction of $\alpha$, namely we can write
\begin{equation}
	\alpha = [0;a_1,\ldots,a_n] =
	\frac{1}{a_1 + \dfrac{\phantom{1}}{\cdots  + \dfrac{1}{a_{n-1} + \dfrac{1}{a_n}}} }  
	\label{eq001}
\end{equation}
where $a_j > 0$, for $j \geq 1$, are positive integer numbers. \\

\begin{figure}[ht]%
	\begin{center}
		\includegraphics[width=14cm]{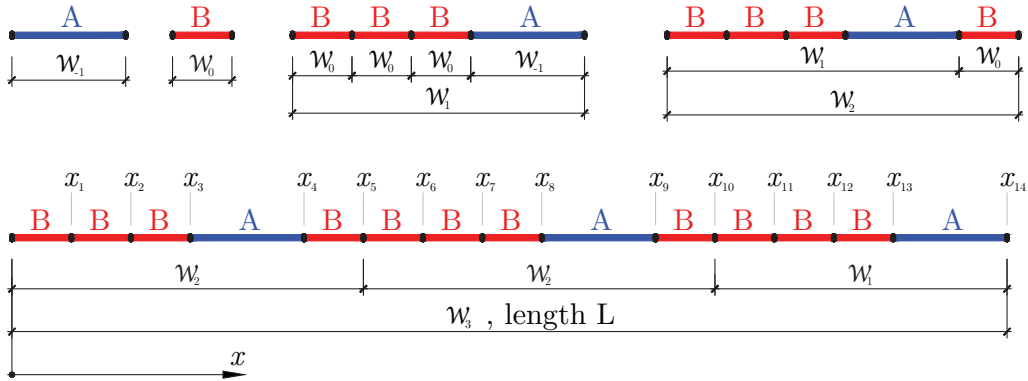} \\	
		\caption{Tiling generation by concatenation for parameter $\alpha = 3/11 = [0;3,1,2]$. Convergents are $\alpha_j = \{1/3, 1/4,3/11\}$. Numerators and denominators of convergents match with the step number of tiles of each type}%
		\label{fig01}%
	\end{center}
\end{figure}

Using the terms of this sequence, a word can be formed from the alphabet $\{A,B\}$ by concatenation. The recursive formula is defined as
\begin{eqnarray}
	\mathcal{W}_j &=& 
	\mathcal{W}_{j-1}^{a_j} \ 
	\mathcal{W}_{j-2}  \ , \quad  1 \leq j \leq n 
	\nonumber \\
	\mathcal{W}_{-1} &=& A \ , \quad \mathcal{W}_{0} = B 
	\label{eq002}
\end{eqnarray}
where both the exponent and the product must be understood as concatenations, for instance $A^3(B^2A) = AAABBA$. The parameter $\alpha$  plays the role of {\em generation parameter} of the quasiperiodic tiling. From the definition given above, if $\alpha$ is a rational number the word $\mathcal{W}_n$ corresponds to the last iteration.  The infinite word emerges as the periodic concatenation of $\mathcal{W}_n$. Otherwise, if $\alpha$ is irrational, then it is known that the sequence $\{a_n\}$ becomes infinite and the associated word has a purely quasiperiodic  pattern given by the limit $\lim_{n\to \infty} \mathcal{W}_n$.  We will refer to the algorithm given by Eq.~\eqref{eq002} as {\em concatenation algorithm}, since the words at each step arise from the concatenation of the previous ones. For instance, for $\alpha = 3/11 = \left[0;3,1,2\right]$ the Fig.~\ref{fig01} shows the different words after each step and the final tiling $\mathcal{W}_3$. 

The final goal is the word $\mathcal{W}_n$ associated to the tiling generated by $\alpha$. However, words of previous steps are somehow approximations. In particular, the number of symbols $A$ and $B$ at each step is given by the sequences $u_j$ and $v_j$ respectively, defined recursively as
\begin{eqnarray}
	u_j &=& a_j \, u_{j-1} + u_{j-2} \ , \qquad u_{-1} =  1 \ , \quad u_{0} = 0  \label{eq003a}  \\
	v_j &=& a_j \, v_{j-1} + v_{j-2} \ , \qquad v_{-1} =  0 \ , \quad v_{0} = 1  \label{eq003b}	
\end{eqnarray} 
Both $u_j$ and $v_j$ are numerator and denominator of the $j$th convergent $\alpha_j = u_j/v_j$\cite{Lazaro-2022a}, with $\alpha = \alpha_n$. Thus
\begin{equation}
	\frac{u_1}{v_1} =  \frac{1}{a_1} \ , \quad
	\frac{u_2}{v_2} = \frac{1}{a_1 + \dfrac{1}{a_2}} \ , \
	\ldots \ , \ \frac{u_n}{v_n} =  \frac{1}{a_1 + \dfrac{\phantom{1}}{\ddots  + \dfrac{1}{a_{n-1} + \dfrac{1}{a_n}}} }    
	\label{eq004}
\end{equation}
Moreover, for two consecutive steps the following identity holds~\cite{Olds-1963}
\begin{equation}
	v_j \, u_{j-1} - u_j \, v_{j-1} = (-1)^j \quad , \ 1 \leq j \leq n
	\label{eq005}
\end{equation}
which leads to the known distance between two consecutive convergents 
$$
\alpha_{j-1} - \alpha_j  =\frac{(-1)^j}{v_j \, v_{j-1}} 
$$
For the tiling associated to the word $\mathcal{W}_j$,  both the total number of points and tiling length are then $N_j  = u_j + v_j$ and $L_{j}  = u_j \, A + v_j \, B$ respectively,  which can also be determined recursively as
\begin{eqnarray}
	N_j &=&  a_j \, N_{j-1} + N_{j-2}
	\ , \quad 1 \leq j \leq n \ , \qquad
	N_{-1} = 1 \ , \quad
	N_{0} = 1\nonumber  \\
	L_{j} &=&  a_j \, L_{j-1} + L_{j-2}
	\ , \quad 1 \leq j \leq n \ , \qquad
	L_{-1} = A \ , \quad
	L_{0} = B 
	\label{eq007} 
\end{eqnarray}
Therefore, the total number of points $N = N_n$ and the final length of the tiling $L = L_n$ are
\begin{equation}
	N = (1 + \alpha) \ v_n \qquad , \quad L = A \, u_n + B \, v_n = (B + \alpha \, A) \, v_n
	\label{eq006}
\end{equation}
\section{Properties of the Structure Factor}

The pattern of points generated by the Sturmian word can be considered as a distribution of local heterogeneities. Thus, the density of the medium generated after $n$ iterations of Eq.~\eqref{eq002} can be written in terms of Dirac-delta functions as
\begin{equation}
	g(x) = \sum_{j=1}^N \delta(x - x_j)
	\label{eq009}
\end{equation}
and its corresponding Fourier series representation leads to 
\begin{equation}
	g(x) = \sum_{m=-\infty}^{\infty} \hat{g}(k_m) \, e^{i k_m \, x}
	\label{eq011}
\end{equation}
where the Fourier coefficients are
\begin{equation}
	\hat{g}(k_m) = \frac{1}{L}\int_{x=0}^{L} \, g(x) \, e^{-i k_m \, x} \, dx\quad , \quad k_m = \frac{2 \pi m }{L} \quad , m = 0, \pm 1, \pm 2, \ldots
	\label{eq010}
\end{equation}
The sequence $\{k_m\}$ represent the reciprocal space positions and according to Eq.~\eqref{eq009} the integral is 
\begin{equation}
	\int_{x=0}^{L} \, g(x) \, e^{-i k \, x} \, dx = \sum_{j=1}^N e^{-i k \, x_j}
	\label{eq012}	
\end{equation}
In general, evaluation of  Eq.~\eqref{eq012} requires in first place the concatenation of the complete word $\mathcal{W}_n$, and secondly the determination the $N$ real space coordinates $x_j, \ 1 \leq j \leq N$. Taking advantage of the recursive formation of the words, we propose a new iterative approach to find the expression of Eq.~\eqref{eq012}, avoiding the computation of the $N$ coordinates $x_j$. This procedure is suitable for quasiperiodic 1D lattices generated by concatenation and it can result useful, especially when handling large systems, which is necessary to simulate aperiodic media. \\

Consider any word $\mathcal{U}$ formed by $U$ symbols taken from the alphabet $\{$A,B$\}$, and real space positions of points given by $\{x_j , \ 1 \leq j \leq U\}$. Let us define
\begin{equation}
	\mathcal{F}\{\mathcal{U};k\} =  \sum_{j=1}^{U} e^{-i k \, x_j}
	\label{eq013}
\end{equation}
As long as there is no room for confusion, we will henceforth refer as Fourier coefficients to those obtained by the above expression \eqref{eq013}. We are interested in evaluating Eq.~\eqref{eq013} for words generated by concatenation. Thus, let us consider $\mathcal{U}$ and $\mathcal{V}$ two arbitrary words formed with symbols taken from the alphabet $\{$A,B$\}$, with lengths $l_u$ and $l_v$ and with a total number of symbols equal to $U$ and $V$, respectively. Let us consider $\{x_j, \ 1 \leq j \leq U \}$ and $\{y_j, \ 1 \leq j \leq V \}$ to be the local positions of tiles for both tilings respectively, verifying that
\begin{equation}
	\mathcal{F}\{\mathcal{U};k\} =  \sum_{j=1}^U e^{-i k \, x_j} \quad , \quad
	\mathcal{F}\{\mathcal{V};k\} =  \sum_{j=1}^V e^{-i k \, y_j}
	\label{eq014}
\end{equation}
Then the word $\mathcal{UV}$ obtained by concatenation has a length $l_u+l_v$ and $U + V $ particles whose coordinates respect to the origin of the concatenated word are
\begin{equation}
	\left\{  x_1, \ldots,x_U, l_u + y_1,\ldots , l_u + y_V \right\}
	\label{eq015}
\end{equation}
The Fourier coefficients of the new tiling is
\begin{equation}
	\mathcal{F}\{\mathcal{UV};k\} = \mathcal{F}\{\mathcal{U};k\} + e^{-ik \, l_u} \, \mathcal{F}\{\mathcal{V};k\}
	\label{eq016}
\end{equation}
Given any integer number $m$ and using the induction principle from this result it is straightforward that
\begin{multline}
	\mathcal{F}\{\mathcal{U}^m;k\} = 
	\left(1 + e^{-i  \, kl_u} + e^{-i \, 2 \, kl_u} + \cdots e^{-i \, (m-1) \, kl_u} \right) \, \mathcal{F}\{\mathcal{U};k\}   \\
	\equiv \mathcal{P}(m,l_u;k)  \, \mathcal{F}\{\mathcal{U};k\}
	\label{eq017}
\end{multline}
where
\begin{equation}
	\mathcal{P}(m,l;k) = 1 + e^{-i  \, k } + e^{-i \, 2 \, kl} + \cdots e^{-i \, (m-1) \, kl} = \frac{1 - e^{-ik\, m \, l}}{1 - e^{-ikl}} 
	\label{eq018}
\end{equation}
stands for the Fourier intensities of a periodic tiling of $m$ particles with separation $l$, that is with coordinates $\{0,l,2l,\cdots,(m-1)l\}$. Making use of the properties given by Eqs. \eqref{eq016} and \eqref{eq017} and denoting by $	\mathcal{H}_j(k) = \mathcal{F}\{\mathcal{W}_j;k\}$ for $1 \leq j \leq n$, it yields
\begin{eqnarray}
	\mathcal{H}_j(k) &=&  \mathcal{F}\{	\mathcal{W}_{j-1}^{a_j} \  \mathcal{W}_{j-2}  ;k\}  \nonumber \\
	&=&     \mathcal{F}\{  	\mathcal{W}_{j-1}^{a_j}     ;k\} + e^{-ik \, L_{j-1}} \, \mathcal{F}\{   \mathcal{W}_{j-2}  ;k\}  \nonumber \\
	&=& \mathcal{P}(a_j, L_{j-1};k) \, 	 \mathcal{H}_{j-1}(k) + e^{-ik \, L_{j-1}} \,  \mathcal{H}_{j-2}(k) \  , \quad 1 \leq j \leq n\label{eq019} \\
	\mathcal{H}_{-1}(k) &=&  e^{-ik  A}	 							  \nonumber \\
	\mathcal{H}_{0}(k) &=&  e^{-ik  B}	 							  \nonumber 
\end{eqnarray}
As this new recursive scheme shows, the Fourier coefficients can be obtained just iterating $n$ times the Eq.~\eqref{eq019} . Thus, it is not necessary to compute the $N$ coordinates of the whole medium, something remarkable from a computational point of view, because in general $N\gg n$. \\

The Fourier coefficients define the properties of the medium in reciprocal space. Media studied in this paper have a discrete distribution of reciprocal space positions $k$ given by $k = 2\pi m /L$, with $m = 0, \pm 1, \pm 2, \ldots$ If $A/B$ is rational, any system obtained by a finite number of interactions $n$ will be periodic having a bounded band of information in the reciprocal space. That is, the distribution of Fourier amplitudes will be periodic. Given a medium of $N$ particles within a length $L$, let us see the period of this distribution. Indeed, the final expression of Fourier coefficients is $\sum \exp\{ikx_j\}$. Assuming $A/B = \theta_A/\theta_B$ is the irreducible fraction of the tiles lengths ratio, then there exist a value of $k$ for which the Fourier coefficients is periodic. That period corresponds to
\begin{equation}
	K_P = \frac{2\pi \theta_B}{B} = \frac{2\pi \theta_A}{A}
	\label{eq020}
\end{equation}
If $A/B$ is irrational, numerator and denominator of the rational approximation $\theta_A/\theta_B$ approaches to infinite making also aperiodic the spectrum in the reciprocal space. It results of interest the fact the period derived in Eq.~\eqref{eq020} is independent of the iterations of Eq. \eqref{eq019}, so that $\mathcal{H}_j(k), \ j \geq 1$ are periodic with period $k = K_P$.  
\begin{figure}[H]%
	\begin{center}
		\begin{tabular}{cc}
			\includegraphics[width=14.0cm]{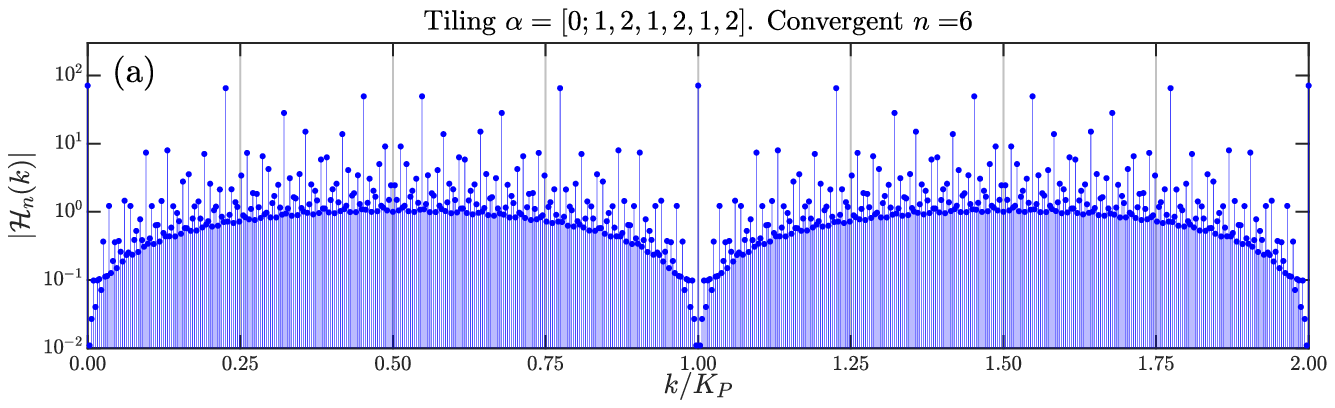}  \\
			\includegraphics[width=14.0cm]{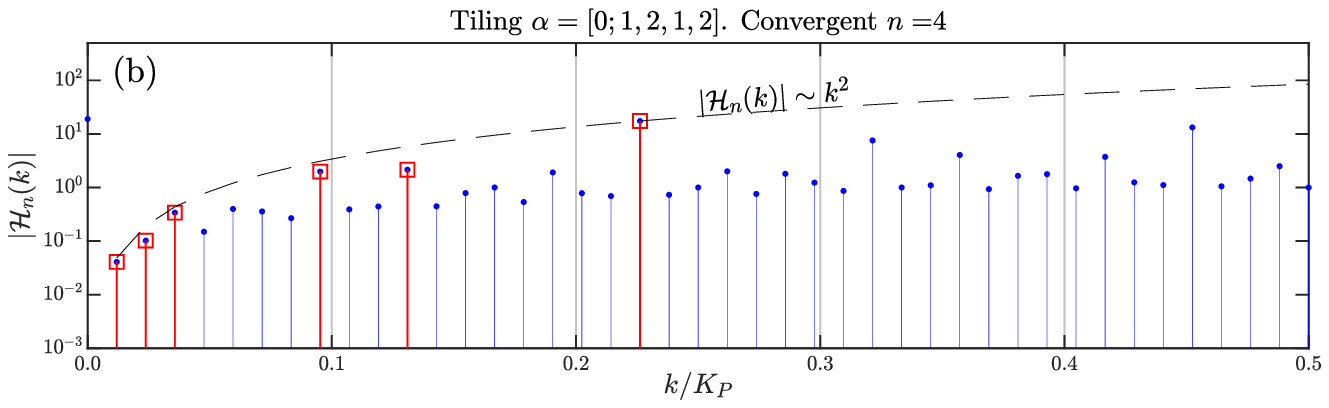}  \\ 
			\includegraphics[width=14.0cm]{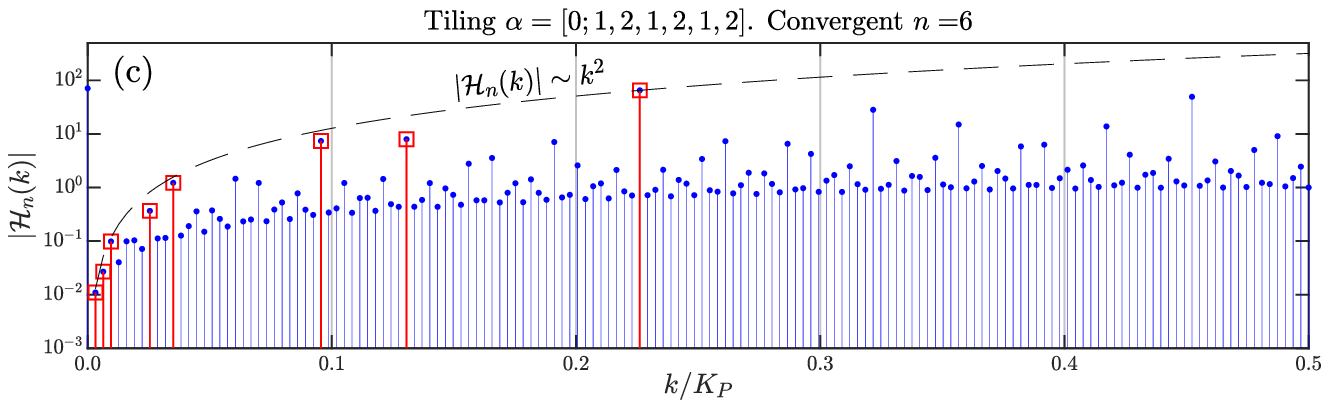}  \\										
			\includegraphics[width=14.0cm]{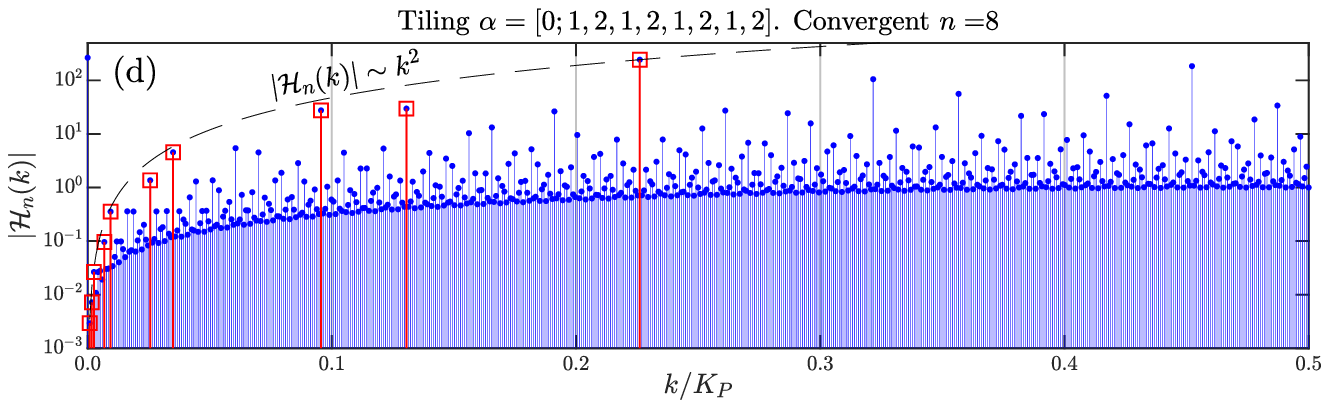}  \\ 
		\end{tabular}
		\caption{Fourier magnitudes for the tiling associated to $\alpha = \sqrt{3}-1 = [0;1,2,1,2,\ldots]$. (a) Fourier magnitudes of the $6$th iteration in the range $0 \leq k \leq 2K_P$. In plots (b), (c) and (d):  Fourier magnitudes of the $4$th, $6$th and $8$th iterations respectivaly in the range $0 \leq k \leq K_P/2$. Blue dots: complete spectrum of Fourier magnitudes. Red squares:  Fourier magnitudes at the sequence of dominant wavenumbers $\{k_\nu\}_{\nu=-1}^n$. Dashed lines: curve proportional to $k^2$}%
		\label{fig02}%
	\end{center}
\end{figure}
Let us illustrate these results graphically with a numerical example. Let us consider the tiling associated to the parameter $\alpha = \sqrt{3}-1 = [0;1,2,1,2,\ldots] = 0.7320508\ldots$ with tiles lengths $A = 1.25$ and $B=1.00$ (units of length), so that $\theta_A = 5$ and $\theta_B=4$. According to Eq.~\eqref{eq020} the spectrum of the Fourier coefficients in the reciprocal space is periodic with period $K_P = 8 \pi$. The magnitudes of the Fourier coefficients corresponding to the $6$th convergent $\alpha_6 = [0;1,2,1,2,1,2]= 0.73170$ are shown in Fig.~\ref{fig02}(a). The range has been extended to $k/K_P=2$ to show the periodicity due to the rational nature of $A/B$. However, within the range $k/K_P=[0,1]$ the self-similarity of the patterns at different scales, typical of quasi-periodic structures, is observed. In Figs.\ref{fig02}(b), \ref{fig02}(c) and \ref{fig02}(d) the image has been enlarged in the range $k/K_P = [0,0.5]$. The three plots represent respectively the Fourier intensities for convergents $n=4$, $n=6$ and $n=8$. It is observed that the limit quasiperiodic tiling present a decreasing intensities as the wavenumbers approach to zero  ($k \to 0$), showing evidences of hyperuniform behavior. Moreover, one particular sequence of Bragg peaks, represented in red, show a more pronounced pattern, proportional to $k^2$. In the next point, the sequence of these characterisitc wavenumbers (drawn in red color in Fig. \ref{fig02}) will be derived.


\section{Sequence of dominant wavenumbers}

Consider the tiling generated by $\alpha = [0;a_1,\ldots,a_n] = u_n/v_n$, with length $L = L_n = A u_n + Bv_n$. For each $j$ with $0 \leq j \leq n$ we can define the  parameter $\sigma_j$ obtained as the  remaining continued fraction after truncation of $\alpha$ up to position $j$, i.e.
\begin{equation}
	\sigma_j = [0;a_{j+1},a_{j+2},\ldots,a_n] =
	\frac{1}{a_{j+1} + \dfrac{ 1}{a_{j+2} + \dfrac{1}{\ddots  + \dfrac{1}{a_{n-1} + \dfrac{1}{a_n}}} }}  
	\label{eq024}
\end{equation}
with $\sigma_0 = \alpha$ and $\sigma_n = 0$. The number $\sigma_j$ plays a key role in the subsequent developments, specially in the computation of the so-called density fluctuations of the tiling. So, it is interesting to derive two different ways of expressing its value, additionally to the continued fraction given by Eq. \eqref{eq024}. First, using the properties of continued fraction~\cite{Olds-1963}, it can be established that
\begin{equation}
	\alpha = [0;a_1,\ldots,a_j,a_{j+1},\ldots,a_n] 
	= \frac{(a_j + \sigma_j) u_{j-1} + u_{j-2}}{(a_j + \sigma_j) v_{j-1} + v_{j-2}} 
	= \frac{u_{j} + \sigma_j u_{j-1}}{v_{j} + \sigma_j v_{j-1}} 
	\label{eq026}
\end{equation}
and solving for $\sigma_j$ 
\begin{equation}
	\sigma_j = - \frac{\alpha - \alpha_j}{\alpha - \alpha_{j-1}} \, \frac{v_j}{v_{j-1}} 
	\label{eq025}
\end{equation}
Secondly, let us express $\sigma_j$ in form of irreducible fraction. For that, we invoke the following decreasing sequence
\begin{equation}
	\xi_j = a_j \, \xi_{j-1} + \xi_{j-2} \ , \quad \xi_{-1} = -v_n \ , \ \xi_0=u_n
	\label{eq032}
\end{equation}
with general term \cite{Lazaro-2022a}
\begin{equation}
	\xi_j = v_n \, v_j (\alpha - \alpha_j)
	\label{eq033}
\end{equation}
Taking into account Eqs \eqref{eq025} \eqref{eq033}, the sequence $\{\xi_j\}$ presents alternating sign and approaches to zero around the limit value $\alpha$. Moreover, in general
\begin{equation}
	\sigma_j = - \frac{\xi_j}{\xi_{j-1}}  \  , \quad  0 \leq j \leq n
	\label{eq034}
\end{equation}
Since the initial values are $ \xi_{-1} = -v_n \ , \ \xi_0=u_n$, then $\xi_j$ will be negative for odd indexes and positive for even ones. Based on this, the sequence $c_j = (-1)^{j-1} \, \xi_{j-1}$ for $0 \leq j \leq n+1$, is formed by positive integer numbers and decreasing  order from $c_0 = v_n$ up to $c_n = 1$ and $c_{n+1}=0$.  The recursive relationship between sequences $\{c_j\}$ and $\{\sigma_j\}$ is then straightforward and given by
\begin{equation}
	c_{j+1} = \sigma_j \, c_j				\ , \quad  0 \leq j \leq n
	\label{eq040}
\end{equation}
Such set of numbers $\{c_j\} $ so formed will be used as basis to build a sequence of wavenumbers associated to the tiling $\alpha_n$. This sequence will be of special importance in the forthcoming developments and it is defined as
\begin{equation}
	k_\nu  = \frac{2\pi}{L_n} \, c_\nu 		\ , \quad 0 \leq \nu \leq n
	\label{eq035}
\end{equation}
Additionally, the term associated to $\nu=-1$ as $k_{-1} = 2\pi N_n/L_n$ will be added at the beginning of the sequence. It should be pointed out that this is a sequence of positive and decreasing numbers, since according to Eq.~\eqref{eq040} each term is obtained by multiplying the previous one by  $\sigma_j$ which is less than the unity. 
In Fig.~\ref{fig02}(b), (c) and (d) the Fourier intensities at these $n+2$ coordinates have been highlighted in red color. As observed, the number of terms of this sequence increases as the corresponding convergent $\alpha_n$ does.  The Fourier intensities at these locations reveal strong periodic patterns in direct space, closely related to the formation of the tiling. Taking a closer look at these wavenumbers for the three systems shown  ($n=4$, $n=6$ and $n=8$), it results clear that their positions fit quite accurately as higher convergents are considered. 
Each column of Table \ref{tab02} shows the numerical results of the $n+2$ wavenumbers $k_\nu/K_P, \ - 1 \leq \nu \leq n$ associated to each generating parameter $\alpha_n$ (including the added wavenumber for $\nu=-1$, introduced above). If, on the other side, we focus on a row, say the $\nu$-th one,  the different values represent how changes the value of $k_\nu$ as the tiling increases. Here is where the results of the table become interesting. Thus, let us consider for instance the values of the wavenumbers $k_0$  ($\nu=0$), which are listed in the second row. As shown above, $c_0 = v_n$, yielding 
\begin{equation}
	k_0 = \left\{   \frac{2\pi v_1}{L_1},  \ \frac{2\pi v_2}{L_2}  , \ldots , \frac{2\pi v_8}{L_8}\right\} = 
	\left\{   \frac{2\pi }{B + A \alpha_1},  \ \frac{2\pi }{B + A \alpha_2}  , \ldots , \frac{2\pi }{B + A \alpha_8}\right\} 
	\label{eq036}
\end{equation}
In general we can denote by $k_0(\alpha_n) = 2\pi / (B + \alpha_n A)$ to the wavenumber for $\nu=0$ calculated for the tiling $\alpha_n = [0;a_1,\ldots,a_n]$. 
From the definition of convergents , we can conclude that if $n \gg 1$, then it is expected that
\begin{equation}
	k_0(\alpha_n) \approx k_0(\alpha_{n+1}) \approx k_0(\alpha_{n+2}) \approx \cdots
	\label{eq037}
\end{equation}
After a quick inspection of the numerical values shown in the second row of Table~\ref{tab02}, we note that the first 4 decimal positions of $k_0$ stabilize from the 6th convergent ($n=6$) onwards. 
\begin{table}[ht]
	\begin{center}
	{  
  \begin{tabular}{|r|rrrrrrrr|}           
  	\hline
  	& \multicolumn{8}{c}{Tiling, $\alpha_n$} \\ 
  	\hline
  										 $\nu$    & $n=1$    & $n=2$    & $n=3$    & $n=4$    & $n=5$    & $n=6$    & $n=7$    & $n=8$ \\ 
  	\hline
  											 $k_{-1}$       & 0.222222 & 0.227273 & 0.225806 & 0.226190 & 0.226087 & 0.226115 & 0.226107 & 0.226109 \\  
  											 $k_0$        & 0.111111 & 0.136364 & 0.129032 & 0.130952 & 0.130435 & 0.130573 & 0.130536 & 0.130546 \\  
  											 $k_1$        & 0.111111 & 0.090909 & 0.096774 & 0.095238 & 0.095652 & 0.095541 & 0.095571 & 0.095563 \\ 
  											 $k_2$        & -        & 0.045455 & 0.032258 & 0.035714 & 0.034783 & 0.035032 & 0.034965 & 0.034983 \\  
  											 $k_3$        & -        & -        & 0.032258 & 0.023810 & 0.026087 & 0.025478 & 0.025641 & 0.025597 \\  
  											 $k_4$        & -        & -        & -        & 0.011905 & 0.008696 & 0.009554 & 0.009324 & 0.009386 \\  
  											  $k_5$        & -        & -        & -        & -        & 0.008696 & 0.006369 & 0.006993 & 0.006826 \\  
  											  $k_6$        & -        & -        & -        & -        & -        & 0.003185 & 0.002331 & 0.002560 \\  
  											  $k_7$        & -        & -        & -        & -        & -        & -        & 0.002331 & 0.001706 \\  
  											  $k_8$        & -        & -        & -        & -        & -        & -        & -        & 0.000853 \\ 
\hline  											  
  	 \end{tabular}%
}
	\end{center}
	\caption{Each row shows the value of $k_\nu/K_P$ for different tilings $\alpha_n$, with $n \geq \nu$, generated by the convergents of the number $[0;1,2,1,2,1,2,\ldots]$. The theoretical results show that after a few iterations, the value of $k_\nu$ stabilizes showing that $k_\nu(\alpha_{n-1}) \approx  k_\nu(\alpha_{n})$ for $n \gg \nu$. }
	\label{tab02}
\end{table}
In general, let us prove that the following expression approximately holds  provided that $n\gg \nu \geq -1$, 
\begin{equation}
	\frac{k_\nu(\alpha_{n-1})}{k_\nu(\alpha_{n})} \approx 1
	\label{eq038}
\end{equation}
Indeed,  from of Eqs.~\eqref{eq040} and \eqref{eq035} it can be established that 

\begin{multline}
	k_\nu(\alpha_n) = \sigma_{\nu-1}(\alpha_n) \, k_{\nu-1}(\alpha_n) = 
	\sigma_{\nu-1}(\alpha_n) \, \sigma_{\nu-2}(\alpha_n)  \, k_{\nu-2}(\alpha_n)  \\ = \cdots =
	\sigma_{\nu-1}(\alpha_n) \, \sigma_{\nu-2}(\alpha_n) \, \cdots \sigma_{0}(\alpha_n) \, k_{0}(\alpha_n)
	\label{eq0038b}
\end{multline}
where the dependence on the associated convergent $\alpha_n$ has been highlighted using notation $k_\nu(\bullet)$ and $\sigma_\nu(\bullet)$. This detail in notation is important at this stage since the quotient of Eq.\eqref{eq038} is the result of evaluating Eq.~\eqref{eq0038b} for two consecutive convergents, $\alpha_{n-1}$ and $\alpha_n$ , indeed
\begin{eqnarray}
	\frac{k_\nu(\alpha_{n-1})}{k_\nu(\alpha_{n})} &=&
	\frac{\sigma_{\nu-1}(\alpha_{n-1})}{\sigma_{\nu-1}(\alpha_{n})} 
	\cdots 
	\frac{\sigma_{1}(\alpha_{n-1})}{\sigma_{1}(\alpha_{n})}
	\frac{\sigma_{0}(\alpha_{n-1})}{\sigma_{0}(\alpha_{n})}
	\frac{k_0(\alpha_{n-1})}{k_0(\alpha_{n})} \nonumber  \\
	&=& 
	\frac{[0;a_{\nu},a_{\nu+1},\ldots,a_{n-1}]}{[0;a_{\nu},a_{\nu+1},\ldots,a_{n}]} 
	\cdots 
	\frac{[0;a_{2},\ldots,a_{n-1}]}{[0;a_{2},\ldots,a_{n}]}      
	\frac{[0;a_{1},\ldots,a_{n-1}]}{[0;a_{1},\ldots,a_{n}]}      
	\frac{B + A \alpha_n}{B + A \alpha_{n-1}} \nonumber \\
	&\approx  & 1 \times \cdots \times 1 \times 1 \approx 1 \ , \quad  n\gg \nu \geq 1
	\label{eq039} 
\end{eqnarray}
The values of each of the fractions $\sigma_j(\alpha_{n-1}) / \sigma_j(\alpha_n)$  for $0 \leq j \leq \nu-1$ are approximately the unity since it is assumed that  $n\gg \nu \geq 1$. Moreover, for $\nu=-1$ and $\nu=0$, it yields too
\begin{eqnarray}
	\frac{k_{-1}(\alpha_{n-1})}{k_{-1}(\alpha_{n})} &=& \frac{2\pi N_{n-1}}{L_{n-1}} \frac{L_{n}}{2 \pi N_{n}} = \frac{1 + \alpha_{n-1}}{1 + \alpha_{n}} \, \frac{B + A\alpha_{n}}{B + A\alpha_{n-1}} 
	\approx 1 \nonumber \\
	\frac{k_{0}(\alpha_{n-1})}{k_{0}(\alpha_{n})} &=& \frac{2\pi v_{n-1} }{L_{n-1}} \frac{L_{n}}{2 \pi \, v_n} = \frac{B + A \alpha_n}{B + A \alpha_{n-1}}
	\approx 1
	\label{eq041}
\end{eqnarray}
These Bragg peaks in reciprocal space have consequences on the behavior of the medium in the long wavelength range. The recursive expression of the Fourier intensities will help to show why this sequence of wavenumbers have dominant magnitudes. Indeed, using Eq.~\eqref{eq019}, the Fourier magnitude associated to the tiling $\alpha_n$ at the wavenumbers $k=k_\nu(\alpha_n) \approx k_\nu(\alpha_{n-1})$ are
\begin{eqnarray}
	\mathcal{H}_n \left[  k_\nu(\alpha_n) \right] &=&
	\mathcal{P} \left[  a_n, L_{n-1};k_\nu(\alpha_n)  \right] \, 	 \mathcal{H}_{n-1} \left[  k_\nu(\alpha_n) \right]  + e^{-i \, k_\nu(\alpha_n) \, L_{n-1}} \,  \mathcal{H}_{n-2}\left[  k_\nu(\alpha_n) \right] 
	\nonumber  \\
	& \approx &
	\mathcal{P} \left[  a_n, L_{n-1};k_\nu(\alpha_{n-1})  \right] \, 	 \mathcal{H}_{n-1} \left[  k_\nu(\alpha_{n-1}) \right]  + e^{-i \, k_\nu(\alpha_{n-1}) \, L_{n-1}} \,  \mathcal{H}_{n-2}\left[  k_\nu(\alpha_{n-1}) \right]   \nonumber \\
	&=& a_n \,	 \mathcal{H}_{n-1} \left[  k_\nu(\alpha_{n-1}) \right]  +  \mathcal{H}_{n-2}\left[  k_\nu(\alpha_{n-1}) \right]  \ , \quad \text{for} \quad n \gg \nu \geq -1
	\\
	\label{eq042}
\end{eqnarray}
where the last step $	\mathcal{P} \left[  a_n, L_{n-1};k_\nu(\alpha_{n-1})  \right] = a_n$ holds because $\mathcal{P}(m,l;k) = m$ when $kl/2\pi$ is a integer number. Following the recursive scheme and, as long as $j$ is sufficiently far from $\nu$, we can approximate  
\begin{eqnarray}
	\mathcal{P} \left[a_j,L_{j-1},k_\nu(\alpha_{j}) \right] & \approx & \mathcal{P} \left[a_j,L_{j-1},k_\nu(\alpha_{j-1}) \right]  = a_j  \nonumber \\
	e^{-i \, k_\nu(\alpha_j) \, L_{j-1}} & \approx & e^{-i \, k_\nu(\alpha_{j-1}) \, L_{j-1}} = 1  \ , \quad n \geq j \gg \nu
	\label{eq043}
\end{eqnarray}
i.e. they take their maximum values. The Fourier intensities $\mathcal{H}_n(k)$ at wavenumbers $k=k_\nu$, with $n \gg \nu$, result maximized with respect to other wavenumbers in their neighborhood. This explains why their intensities are several orders of magnitude larger than the rest. Furthermore,  in Fig. ~\ref{fig02}, it can be observed that the law of the form $\left| \mathcal{H}_n(k) \right| \sim k^2$, seems to fit more accurately at the Bragg peaks of the sequence $\{k_\nu\}$. \\

The approximation of Eqs.\eqref{eq039} and \eqref{eq041} gets worse as the value of $k_\nu(\alpha_{n-1})$ differs from $k_\nu(\alpha_{n})$, which becomes  evident as $\nu$ gets closer to $n$. However, recall we can make $n$ as large as we want for a pure quasiperiodic tiling generated by  an infinite continued fraction. Thus, the Fourier intensities at these Bragg peaks at the sequence $\{k_\nu\}$ decays towards $k\to 0$ following a pattern stronger than the linear one. In fact, we will see in the next section that, under certain assumptions tested in previous works we can predict analytically the Fourier intensities in this sequence of wavenumbers, which will be called  \emph{sequence of dominant wavenumbers}.

\section{Hyperuniformity exponent}

It is well known that the order of hyperuniformity of a medium in reciprocal space is closely related to the limiting values of density fluctuations in the physical space. In the particular case of quasiperiodic media generated by substitution, it has been observed that the hyperuniformity and the limit density fluctuations are close related~\cite{Torquato-2018a} and in turn these latter are proportional to the ratio of the two eigenvalues of the substitution matrix. Our goal is to extend these results to the family of quasiperiodic systems generated by concatenation of words using a continued fraction, as shown in Eq.~\eqref{eq002}. Let us consider a tiling based on the convergent $\alpha = [0;a_1\ldots,a_n,]$, with $n \gg 1$. As the words $\{\mathcal{W}_j, \ 1 \leq j \leq n\}$ are generated, both the number of tiles $N_j$ and the length of the tiling $L_j$ become larger. The density of  points associated to the $j$th convergent $\alpha_j = [0;a_1,\ldots,a_j] = u_j / v_j$ can be determined as $\rho_j = N_j / L_j$. From Eqs.~\eqref{eq007} and \eqref{eq006} and after $j$ iterations, the density of points yields
\begin{equation}
	\rho_j = \frac{N_j}{L_j} = \frac{u_j + v_j}{Au_j + Bv_j}  = \frac{1 + \alpha_j }{B + \alpha_j  A}
	\label{eq021}
\end{equation}
Denoting by $\bar{\rho} = N/L = (1+\alpha) / (B + \alpha A)$ to the limit density of tile vertices, then we can writhe $\rho_j =  \bar{\rho} + \delta \rho_j$, where $\delta \rho_j$ stands for the deviations respect to $\bar{\rho}$ and they are given by 
\begin{equation}
	\delta \rho_j = \frac{(A-B) (\alpha - \alpha_j )}{(B + \alpha_j  A) (B + \alpha A)} \quad , \ 1 \leq j \leq n
	\label{eq022}
\end{equation}
This relationship exhibits the decreasing amplitudes of density fluctuations for large scales, characteristic of hyperuniform structures. 
From Eq.~\eqref{eq022}, the ratio between density fluctuations for two consecutive iterations yields
\begin{equation}
	\frac{	\delta \rho_j}{	\delta \rho_{j-1}} = \frac{\alpha - \alpha_j }{\alpha - \alpha_{j-1} } \  \frac{B + A \, \alpha_{j-1} }{B + A \alpha_{j}}
	\label{eq023}
\end{equation}
This expression reveals the close relationship between the ratio of density fluctuations and the parameter $\sigma_j$  introduced in the previous section. Furthermore, using Eq. \eqref{eq025} we can rewrite the above equation as
\begin{equation}
	\frac{	\delta \rho_j}{	\delta \rho_{j-1}} = - \sigma_j \, \tau_j
	\label{eq044}
\end{equation}
where the new parameter $\tau_j = L_{j-1}/L_j$ denotes the relationship between tiling lengths at two consecutive iterations. As shown in Eq.~\eqref{eq007}, the sequence of the tiling lengths obeys the recursive scheme given by $L_j = a_j L_{j-1} + L_{j-2}$. Therefore, the parameter $\tau_j$ can be expressed in other form as
\begin{equation}
	\tau_j = \frac{L_{j-1}}{L_j} = 
	\frac{1}{a_{j} + \dfrac{ 1}{a_{j-1} + \dfrac{1}{\ddots  + \dfrac{1}{a_{1} + \dfrac{\theta_A}{\theta_B}}} }}  = 
	[0;a_{j},a_{j-1},\ldots,a_1 + \theta_A/\theta_B]
	\label{eq047}
\end{equation}

Eq. \eqref{eq044} reveals that density fluctuations decay following an exponential-type law and alternating the corresponding sign around the average density. It is known that one-dimensional quasiperiodic media generated by substitution rules exhibit density fluctuations that depend on the eigenvalues of the substitution matrix \cite{Torquato-2018a}. Moreover, in such media it has been found  \cite{Baake-2018,Godreche-1990} that the Fourier intensities are scaled under the same pattern as the density fluctuations. Aperiodic tilings studied in this paper are built by means of word concatenation, governed by a generic continued fraction $[0;a_1,\ldots,a_n]$. Thus, each new word depends on a new number $a_j$ given by the continued fraction, making them of special nature. 
After the definition of the sequence of dominant wavenumbers, see Eqs.~\eqref{eq040} and \eqref{eq035}, and considering the derived expression for the density fluctuations ratio in Eq.~\eqref{eq044}, two major facts have been identified
\begin{itemize}
	\item [(i)] According to Eq.~\eqref{eq042}, the Fourier intensities are maximized at the dominant wavenumber sequence $k_{\nu+1} = \sigma_{\nu} \, k_\nu$
	\item [(ii)] Ratio of two consecutive density fluctuations is proportional to the ratio between two consecutive dominant wavenumbers, i.e. $\frac{	\delta \rho_{\nu}}{	\delta \rho_{\nu-1}} = - \sigma_\nu \, \tau_\nu$
\end{itemize}
For the purposes of this section, we can ignore the subscript $n$ since other convergents will not be of interest. Thus, for convenience in notation, let us denote as $H(k) = \left| \mathcal{H}_n(k)\right|$ to the magnitude of the Fourier intensity of tiling generated by $\alpha = [0,a_1,\ldots,a_n]$ at wavenumber $k$. Assuming the hypothesis that Fourier intensities scales as the density fluctuations, we can establish the following relationship for each pair of consecutive wavenumbers within the sequence $\{k_\nu\}_{\nu=0}^{n-1}$
\begin{equation}
	H(\sigma_\nu \, k_\nu ) =  \left| \frac{	\delta \rho_\nu}{	\delta \rho_{\nu-1}} \right|  \cdot	{H}(k_{\nu})
	\label{eq046}
\end{equation}
Assuming a power law for the Fourier magnitudes, we find that
\begin{equation}
	\frac{H(k_{\nu+1})}{H(k_\nu)} = \left( \frac{ k_{\nu+1} }{ k_{\nu} } \right)^{1 + \log \tau_\nu / \log \sigma_\nu} \quad , \quad 0 \leq \nu \leq n-1
	\label{eq045}
\end{equation}
where both $\sigma_\nu$ and $\tau_\nu$ can be written as the  continued fractions
\begin{eqnarray}
	\tau_\nu &=&	[0;a_{\nu},a_{\nu-1},\ldots,a_1 + \theta_A/\theta_B] \nonumber \\
	\sigma_\nu &=& [0;a_{\nu+1},a_{\nu+2},\ldots,a_n] 
	\label{eq048}
\end{eqnarray}
Therefore, the structure factor decays with wavenumbers according the law
\begin{equation}
	\frac{S(k_{\nu+1})}{S(k_\nu)} = \left( \frac{ k_{\nu+1} }{ k_{\nu} } \right)^{2 + 2\log \tau_\nu / \log \sigma_\nu} \quad , \quad 0 \leq \nu \leq n-1
	\label{eq045b}
\end{equation}
As Eqs. \eqref{eq045} and \eqref{eq045b} show, the scaling factor in the Fourier intensities is variable for each step depending on the continued fraction sequence $\{a_j, \ 1 \leq j \leq n\}$. Thus, new higher terms of the sequence $\{a_n\}$ provide information on successive scales in the large wavelength range, or in other words, in the different small scales in the reciprocal space, around $k \to 0$. Let us illustrate the proposed model of Eq.~\eqref{eq045} with an example. \\

\begin{figure}[H]%
	\begin{center}
		\includegraphics[width=16cm]{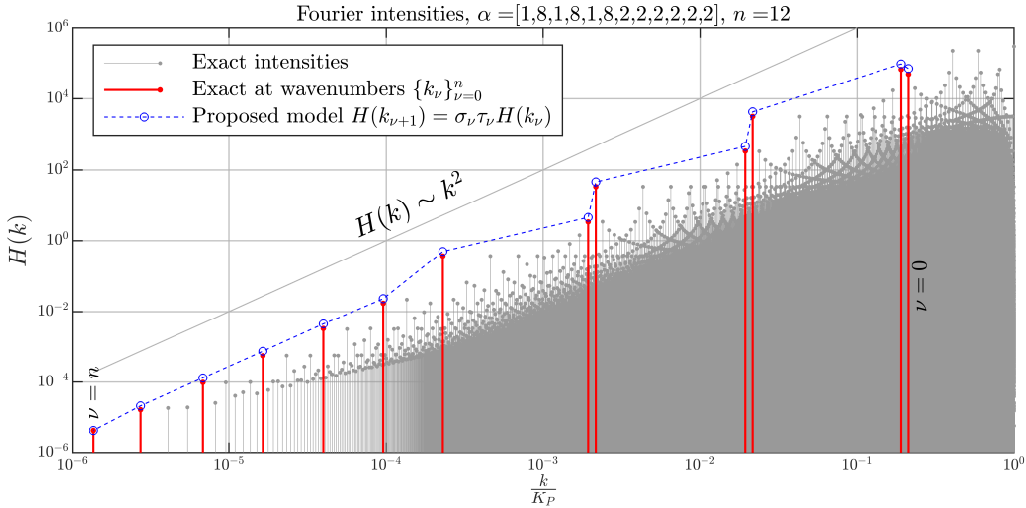} \\	
		\caption{Fourier intensities for a quasiperiodic tiling generated by $\alpha = [0;1,8,1,8,1,8,2,2,2,2,2]$. In red color: intensities at the wavenumbers of the dominant sequence $k_\nu = 2\pi c_\nu/L, \ 0 \leq \nu \leq n$. In blue color: Proposed model based on density fluctuations}%
		\label{fig03}%
	\end{center}
\end{figure}

Let us consider for that the tiling generated by 
$$
\alpha = [0;1,8,1,8,1,8,2,2,2,2,2,\ldots] \approx 0.8989789\ldots
$$
The Fourier intensities can be determined using the recursive procedure proposed in Eq.~\eqref{eq019} for each wavenumber $k = 2\pi m/L, \ m = 0,\pm1,\pm 2,\ldots$. They have been plotted in Fig.~\ref{fig03}, highlighting in red color the Bragg peaks at the dominant wavenumbers $k_\nu = 2\pi c_\nu/L, \ 0 \leq \nu \leq n$, defined in Eqs.~\eqref{eq035}. The first term of the sequence, for $\nu=0$ is also the highest one, with value $k_0 = 2\pi v_n/L \approx 2\pi/(B+A\alpha)$. As the index $\nu$ increases in the range $0 \leq \nu \leq n$, the value of $k_\nu$ decays up to the last (and lowest) value $k_n = 2\pi/L$. The subsequent Fourier intensities $H(k_\nu)$ from $\nu = n-1$ up to $\nu=0$ can be obtained recursively from the previous ones by means of proposed approach of Eq.~\eqref{eq045}, starting from $H(k_n)$. These values are shown with blue-dashed line shown in Fig.~\ref{fig03}. The proposed method satisfactorily fits the exact results of the spectrum in the reciprocal space at the coordinates given by the dominant wavenumbers $\{k_\nu\}_{\nu=0}^{n}$. In order to obtain the Fig.~\ref{fig03}, the 12th  approximant of $\alpha$ $(n=12)$ has been considered. As known, the sequence $k_{\nu}$ follows the recursive scheme $k_{\nu+1} = \sigma_{\nu} \, k_{\nu}$. 
\begin{table}[ht]
	\begin{center}
	{  \begin{tabular}{lrrrrrr}           
	\\
	\cline{2-7}
													& $\sigma_0$    & $\sigma_1$    & $\sigma_2$    & $\sigma_3$    & $\sigma_4$    & $\sigma_5$    \\ 
	\hline \\
	Rational form, $c_{\nu+1}/c_{\nu}$      & $\frac{140078}{155819}$ &  $\frac{15741}{140078}$ & $\frac{14150}{15741}$ 
													& $\frac{1591}{14150}$ & $\frac{1422}{1591}$  & $\frac{169}{1422}$  \\ \\
    Decimal form, $\sigma_\nu$        			& 0.8989 & 0.1124  & 0.8989  & 0.1124 & 0.8928 & 0.1188 \\ \\ \\
	\cline{2-7}
													& $\sigma_6$    & $\sigma_7$    & $\sigma_8$    & $\sigma_9$    & $\sigma_{10}$    & $\sigma_{11}$    \\ 
	\hline \\
	Rational form,	$c_{\nu+1}/c_{\nu}$      & $\frac{70}{169}$ &  $\frac{29}{70}$ & $\frac{12}{29}$ 
													& $\frac{5}{12}$ & $\frac{2}{5}$  & $\frac{1}{2}$  \\ \\
	Decimal form, $\sigma_\nu$        			& 0.4142 & 0.4143  & 0.4138  & 0.4167 & 0.4000 & 0.5000 \\ \\					
	\hline						  
\end{tabular}%

}
	\end{center}
	\caption{Values of the parameter $\sigma_\nu = c_{\nu+1}/c_{\nu}$, for $0 \leq \nu \leq 11$ both in rational and decimal form, obtained from the continued fraction $\alpha = [0;1,8,1,8,1,8,2,2,2,2,2]$. The last value for $n=12$ is $\sigma_{12}=0$}
	\label{tab01}
\end{table}
In the particular case of $\alpha = [0;1,8,1,8,1,8,2,2,2,2,2]$, the sequence of numbers $\{\sigma_{\nu}\}$ have been listed in Table \ref{tab01}, both in rational and in decimal form. The generator parameter $\alpha$ in this example has been carefully chosen with the first six terms alternating between 1 and 8 and the second six terms constant and equal to 2. This choice allows us to show the interesting property demonstrated in Eq. \eqref{eq045}:  the Fourier intensities  locally behave according to the pattern of the sequence $\{a_j\}_{j=1}^{n}$. Indeed, the sequence of dominant wavenumbers is obtained from the values  $\sigma_{\nu}$ listed in Table \ref{tab01}. Thus, the first of them (ordered from highest to lowest are)
\begin{equation}
	k_1 = 0.8989 \, k_0 \ , \quad  k_2 = 0.1124  \, k_1 \ , \quad  k_3 = 0.8989 \, k_2   \ , \quad  k_4 = 0.1124 \, k_3 \ , \ldots 
	\label{eq075} 
\end{equation}
It follows that $k_0$ and $k_1$ are quite close to each other, but $k_1$ and $k_2$ are far apart. These distances between the wavenumbers are the reflection in the reciprocal space of the jumps between the values 1 and 8 in the sequence. On the other hand, when we evaluate the wavenumbers from $k_7$ onwards we find
\begin{equation}
	k_7 = 0.4142 \, k_6  \  , \quad k_8 = 0.4143k_7 \  , \quad k_9 = 0.4128 \, k_8  \  , \quad k_{10} = 0.4167 \, k_9 \ , \ldots
	\label{eq076} 
\end{equation}
i.e., from $\nu\geq 7$, the wavenumbers are equidistant (in logarithmic scale), reflecting the constant behavior of the sequence as $a_7=a_8 = \cdots =2$. The behavior described here can be clearly seen in the red-colored coordinates of the dominant wavenumbers in Fig. \ref{fig03}.

However, if we look at the overall order of decay of the Fourier intensities as $k \to 0$ in Fig.~\ref{fig03}, on average it results very similar to a quadratic law. Let us  see in the following sections some results which demonstrate indeed that the quasiperiodic tilings generated by continued fractions are hyperuniform with exponent equal to 3, i.e. $S(k) \sim k^4$. Rigorous proofs will be carried out for the cases of metallic means and for periodic continued fractions. Furthermore, we will prove that for any other system, the order of decay of the structure factor intensities along the whole sequence $\{k_\nu\}_{\nu=0}^{n}$ is asymptotically a 4th order power law.

\subsection{Metallic means: $\alpha = [0;a,a,a,\ldots]$}

This case collects the behavior of generalized Fibonacci-type quasiperiodic media that could also be simulated using the substitution rule $A \to B$, $B \to B^a A$. The order of hyperuniformity has been studied in refs. \cite{Torquato-2017, Torquato-2018a} resulting in a structure factor decaying with the power law $S(k) \sim k^4$, meaning that the Fourier intensities decay strictly quadratically. Let us see that this result can be derived from the model presented in this work. As known~\cite{Macia-2009} the metallic means,  represented by the continued fraction $\alpha = [0;a,a,a,\ldots]$, are solutions of the quadratic equation $\alpha^2 = 1 + a \, \alpha$. The two first values of the sequences $\sigma_\nu$ and $\tau_\nu$ are 
$	\sigma_{0} = \alpha  , \  \tau_0 = \theta_A/\theta_B$. Since $a_j$ is constant, after several steps $\sigma_\nu$ and $\tau_\nu$ become approximately equal. Assuming then $n \gg \nu \gg 1 $ it yields
\begin{equation}
	\sigma_{\nu} = [0;a,a,\ldots] =\alpha \quad , \quad \tau_\nu = [0;a,a,\ldots,a+\theta_A/\theta_B] \approx \alpha
\end{equation}
so that the relationship $H(\sigma_\nu k_\nu) = \sigma_\nu \, \tau_\nu \, H(k_\nu)$ can be approximated by
\begin{equation}
	H(\alpha k_\nu) = \alpha^2 \, H(k_\nu) 
	\label{eq050}
\end{equation}
Therefore, the Fourier intensities at the wavenumbers $k = k_\nu$ can be simulated according to the quadratic law $H(k) \sim k^2$ ($k \to 0$). \\

\begin{figure}[h]%
	\begin{flushleft}
		\begin{tabular}{cc}
			\includegraphics[width=8cm]{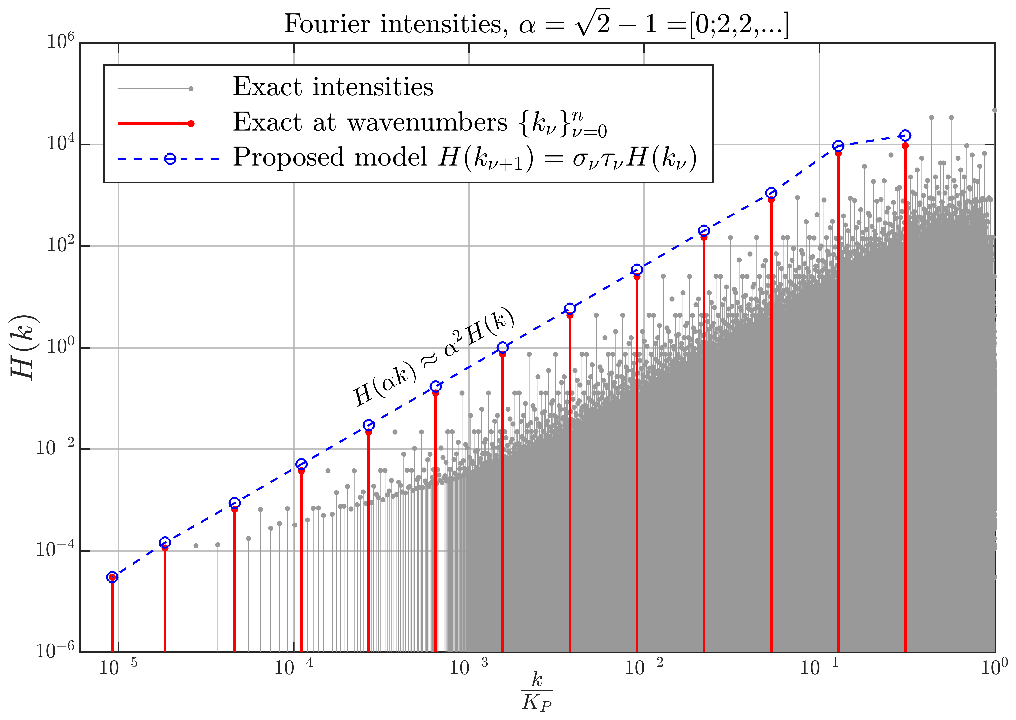}  &
			\includegraphics[width=8cm]{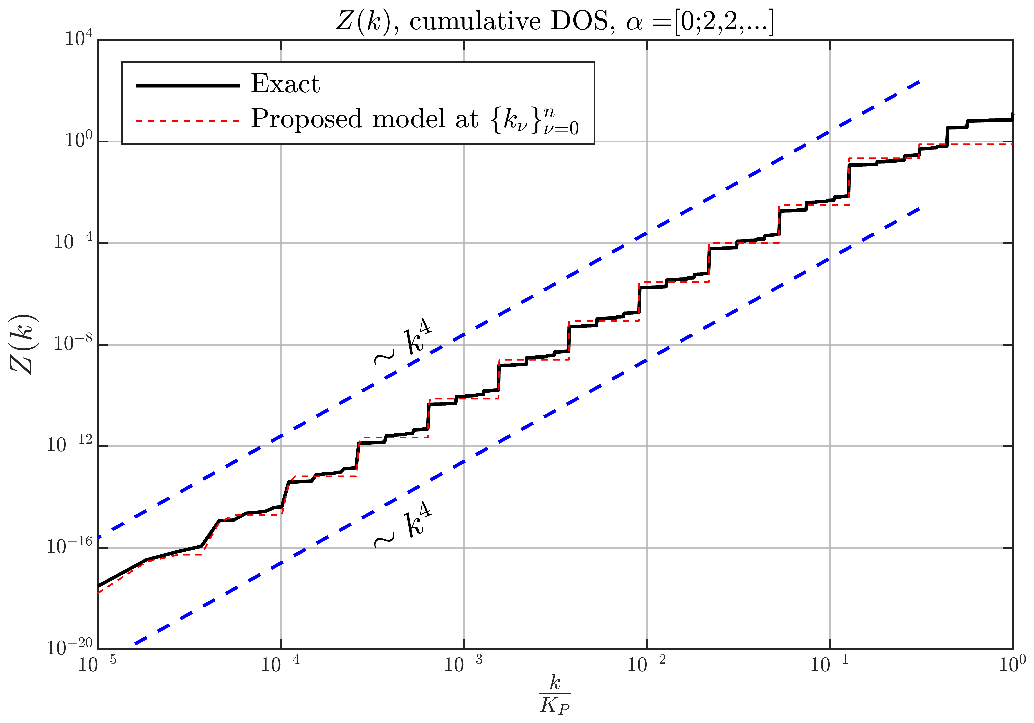}  \\ 
			\includegraphics[width=8cm]{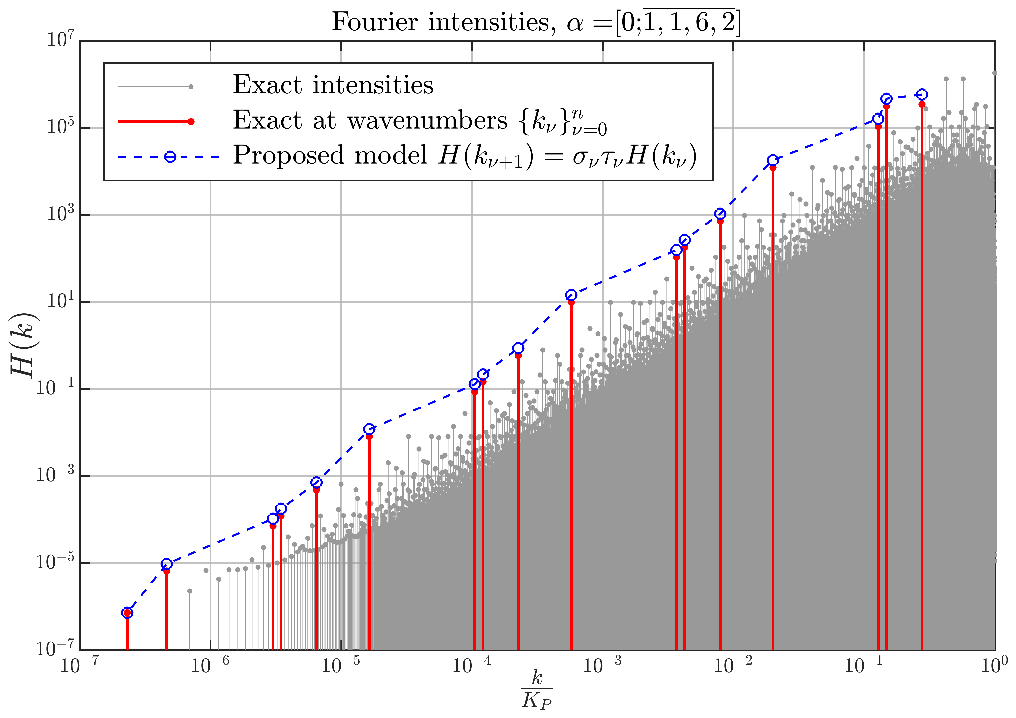}  &										
			\includegraphics[width=8cm]{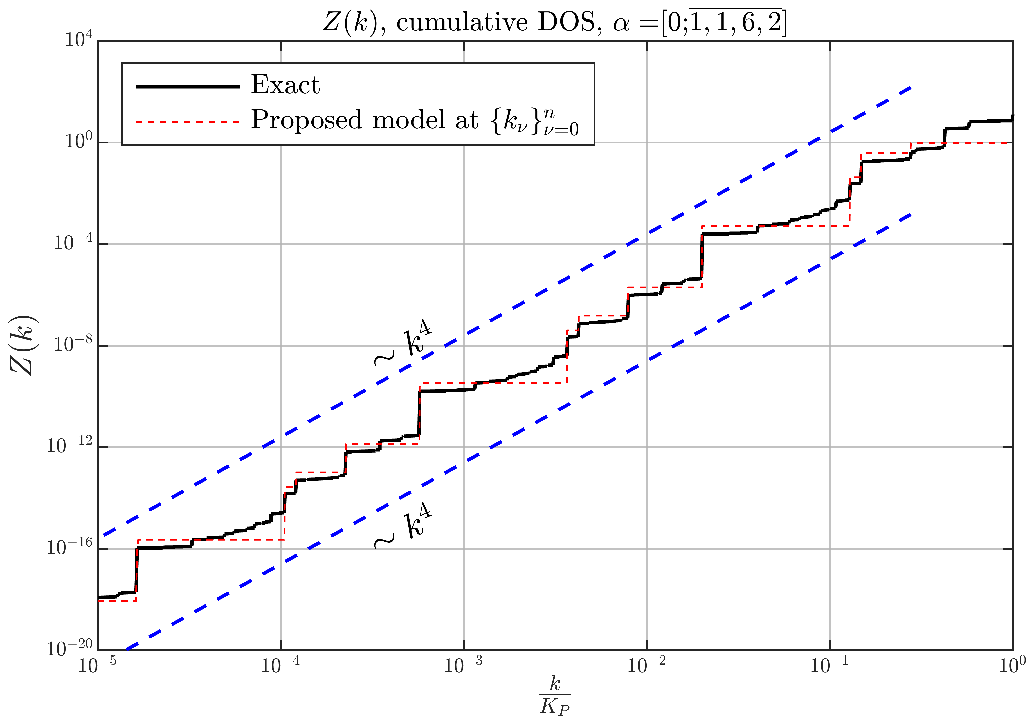}  \\ 
		\end{tabular}
		\caption{Fourier intensities (left) and cumulative intensity function (right) for two quasiperiodic tilings generated by the silver mean $\alpha = \sqrt{2}-1 = [0;2,2,2,\ldots]$ (top) and by the periodic continued fraction $\alpha = [0;\overline{1,1,6,2}]$. In left plots: gray peaks represent the exact Fourier intensities at wavenumbers $k = 2 \pi m/L, \ m=0,1,2,\ldots$, red peaks represent the exact Fourier intensities at dominant wavenumbers $k_\nu = 2 \pi c_\nu/L$, defined in Eq.~\eqref{eq035}, blue dots: Fourier magnitudes at sequence of wavenumbers $\{k_\nu\}$ obtained by the approximate model. In right plots: black line represents the exact cumulative intensity function, red line represents the cumulative function but obtained from the approximated Bragg peaks at the dominant wavenumbers. Blue dashed line represents the 4th power order envelope curves}%
		\label{fig04}%
	\end{flushleft}
\end{figure}

It is straightforward that this behavior towards  $k \to 0$ (long wavelength range) is governed by the latest values of  sequence $\{a_j\}$ (those with highest values of the index $j$) which, from the definition of the sequence $\{k_\nu\}$, are associated to the lowest values of the wavenumbers. Therefore, it is clear that the Fourier intensities will also decay quadratically for tilings generated by continued fractions of the form $\alpha = [0;d_1,\ldots,d_m,a,a,a,\ldots]$, such as the one shown in Fig.~\ref{fig03}. In Fig.~\ref{fig04} (top), both the Fourier intensities and the cumulative function $Z(k)$, defined in Eq.~\eqref{eq000b}, have been represented for the case $\alpha = \sqrt{2}-1 = [0;2,2,2\ldots]$. \\

Since $S(k)$ is formed by a set of singular peaks, it should not be induced that the order of the cumulative intensity function is of one order higher. On the contrary, in these cases it turns out that both $S(k)$ and $Z(k)$ share the same exponent \cite{Torquato-2018a}. 
In fact, numerical simulations carried out in this paper shows that, as for the Fibonacci projection cases \cite{Torquato-2017,Torquato-2018a}, the scaling of Fourier peaks and their locations produces the cumulative function $Z(k)$ to scale under the same power-exponent as  $S(k)$, showing that this property also holds for quasiperiodic tilings generated by continued fractions  (see  Fig.~\ref{fig04} ). Therefore, it follows that $Z(k) \sim k^4$, which, according to Eq.~\eqref{eq00d}, immediately leads to an exponent of hyperuniformity $\gamma=3$. The discrete nature of the spectrum makes the function $Z(k)$ to behave like a cumulative step-wise function as shown in Fig.~\ref{fig04}(right).

\subsection{Periodic continued fractions: $\alpha = [0;\overline{a_1,\ldots,a_p}]$}

Fig. \ref{fig03}  shows that the presence of certain repeating pattern in the sequence $\{a_j\}$ makes the Fourier intensities also reveals periodicity as smaller wavenumbers $k$ are considered. Still, the decay rate of the intensities seems to be quadratic globally, although locally they can be either greater or lower than 2. In this section, it will be proved that, indeed, the global decay order of the Fourier intensities for periodic continued fractions of the form $\alpha = [0;\overline{a_1,\ldots,a_p}]$ is exactly 2. The overline notation represents repetition, i.e.
\begin{equation}
	\alpha = [0;\overline{a_1,\ldots,a_p}] = [0;a_1,\ldots,a_p,a_1,\ldots,a_p,\ldots]
	\label{eq071}
\end{equation}
Considering $ n \gg \nu \gg p$, we can introduce the following values
\begin{equation}
	\hat{	\sigma} = \sigma_{\nu+1} \cdot \sigma_{\nu+2} \cdots \sigma_{\nu+p} \ , \quad
	\hat{\tau} = \tau_{\nu+1} \cdot \tau_{\nu+2} \cdots \tau_{\nu+p}
	\label{eq072}
\end{equation}
Since the assumed model is multiplicative for both the Fourier intensities and the  sequence $\{k_\nu\}$, the value $\hat{ \sigma} = k_{\nu+p+1} / k_{\nu+1}$ represents the global jump between the two wavenumbers $k_{\nu+1}$ and $k_{\nu+p+1}$, separated $p$ steps from each other. Due to the periodicity of the continued fraction, the value of $\hat{\sigma}$ is independent of the value of $\nu$ considered. With a sufficiently high value of $\nu$ fixed, from $k_{\nu+1}$ onwards, the relationship between the Fourier intensities over $p$ steps is
\begin{multline}
	H(k_{\nu+p+1}) = \sigma_{\nu+p} \, \tau_{\nu+p} \,   	H(k_{\nu+p}) \\  
	= \left[ \sigma_{\nu+p}  \cdots  \sigma_{\nu+1} \right] \,   
	\left[ \tau_{\nu+p}  \cdots  \tau_{\nu+1} \right] 	H(k_{\nu+1}) = \hat{	\sigma} \hat{\tau} \,  H(k_{\nu+1})
	\label{eq051}
\end{multline}
Since the relationship between the wavenumbers is 
$$k_{\nu+p+1} = \sigma_{\nu+1} \cdots \sigma_{\nu+p}  \, k_{\nu+1} \equiv \hat{\sigma}  k_{\nu+1} \ ,$$ 
we can write that
\begin{equation}
	H(\hat{\sigma} k_{\nu+1}) = \hat{	\sigma} \hat{\tau} \,  H(k_{\nu+1})
	\label{eq052}
\end{equation}
Let us now see that the two values $\hat{\sigma}$ and $\hat{\tau}$ are approximately equal provided that $n \gg \nu \gg 1$, where $n$ represents the total size of the sequence $\{a_j\}$. In fact, in order to achieve this, we will determine both  $\hat{\sigma}$ and $\hat{\tau}$  separately, finding for them more compact expressions. From Eq.~\eqref{eq034}, it is
\begin{eqnarray}
	\hat{\sigma} &=& \sigma_{\nu+1} \cdot \sigma_{\nu+2} \cdots \sigma_{\nu+p}  \nonumber \\
	&=&   \left( - \frac{\xi_{\nu+1}}{\xi_{\nu}} \right) \, \left( - \frac{\xi_{\nu+2}}{\xi_{\nu+1}} \right) \cdots \left( - \frac{\xi_{\nu+p}}{\xi_{\nu+p-1}} \right) \nonumber \\
	&=& (-1)^p \, \frac{\xi_{\nu+p}}{\xi_{\nu}} \label{eq053}
\end{eqnarray}
Now, since $\hat{\sigma}$ does not depend on $\nu$, we can choose any index $\nu$ to obtain its value. In particular, it is of interest to take $\nu=0$, for which Eq.~\eqref{eq053} is found to be  $\hat{\sigma} = (-1)^p \xi_p / \xi_0$. Using the expression from Eq. \eqref{eq033}, we can calculate $\xi_p$ as
\begin{equation}
	\xi_p = v_n \, v_p  \left(\alpha - \frac{u_p}{v_p}\right)
	\label{eq055}
\end{equation}
where as known $\alpha = u_n/v_n$ and  $u_p/v_p = [0;a_1,\ldots,a_p]$ denotes the $p$th convergent. Additionally, $\xi_0 = u_n$, so that the value of $\hat{\sigma}$ can be expressed finally as
\begin{equation}
	\hat{\sigma} = (-1)^p \, \frac{\xi_{p}}{\xi_{0}} = (-1)^p \, \frac{v_n \, v_p}{u_n} \left(\alpha - \frac{u_p}{v_p}\right) = 
	(-1)^p \left(v_p - \frac{u_p}{\alpha}\right) 
	\label{eq054}
\end{equation}
On the other hand, it turns out that the expression for $\hat{\tau}$ given in Eq.~\eqref{eq072} can also be meaningfully abbreviated. Each $\tau_j, \ \nu+1 \leq j \leq \nu+p$ is defined as the ratio between two consecutive tiling lengths, that is 
$\tau_j = L_{j-1}/L_j = [0; a_j,a_{j-1},\ldots,a_1+\theta_A/\theta_B]$, hence they all are finite continued fractions. However, since it is assumed that $\nu \gg p$, then for $\nu+1 \leq j \leq \nu+p$, $\tau_j$ can be approximated as
\begin{multline}
	\tau_j =  [0; a_j,a_{j-1},\ldots,a_1, \ a_p,\ldots,a_1, \ldots,a_p,\ldots,a_1+\theta_A/\theta_B] \\
	\approx 
	[0; a_j,a_{j-1},\ldots,a_1, \overline{a_p,\ldots,a_1}]
	\label{eq073}
\end{multline}
where the last expression is an infinite continued fraction. Therefore, the above assumption allow us to write each $\tau_j$ as function of  the parameter $\beta = [0;\overline{a_p,a_{p-1},\ldots,a_1}]$ obtained from $\alpha$ by reversing the period. Indeed, 
\begin{eqnarray}
	\hat{\tau} &=& \tau_{\nu+1} \cdot \tau_{\nu+2} \cdots \tau_{\nu+p} \nonumber \\
	&\approx & [0;a_1,\overline{a_p,\ldots,a_1}]   
	[0;a_2,a_1,\overline{a_p,\ldots,a_1}]  
	\cdots [0;a_{p-1},\ldots,a_1,\overline{a_p,\ldots,a_1}]  \cdot
	[0;\overline{a_p,\ldots,a_1}]  \quad  \nonumber \\
	&=& 	\frac{1}{a_{1} + \beta} \cdot \frac{1}{a_{2} +  \dfrac{1}{a_1 + \beta} } \cdot \cdots \cdot
	\frac{1}{a_{p-1} + \dfrac{ 1}{a_{p-2} + \dfrac{1}{\ddots  + \dfrac{1}{a_{1} + \beta }} }}  \cdot \beta
	\label{eq061}
\end{eqnarray}
The above expression shows that $\hat{\tau}$ is constant and independent of $\nu$ when considering values $\nu \gg p$ something that will be used later. Using the definition given by $\tau_j = L_{j-1}/L_j$, see Eq.~\eqref{eq047}, one can simplify the value of $\hat{	\tau}$ as 
\begin{equation}
	\hat{\tau} = \tau_{\nu+1} \cdot \tau_{\nu+2} \cdots \tau_{\nu+p} = 
	\frac{L_\nu}{L_{\nu+1}} \cdot \frac{L_{\nu+1}}{L_{\nu+2}} \cdots \frac{L_{\nu+p-1}}{L_{\nu+p}} = \frac{L_{\nu}}{L_{\nu+p}}
	\label{eq056}  
\end{equation}
The tilings lengths obey the characteristic recursive sequence $L_j = a_j \, L_{j-1} + L_{j-2}$ as shown in Eq.~\eqref{eq007}. Since, as shown above in Eq.~\eqref{eq061}, $\hat{\tau}$ is independent of $\nu$, we can match the latter with a value $\nu$ multiple of the period, i.e., $\nu = m\, p$, where $m$ is some large natural number. The tiling length in step $\nu+p$ can then be expressed in terms of the lengths of the previous steps up to step $L_{\nu}$. Following the sequence and using the properties of the corresponding sequences \cite{Lazaro-2022a}, we have
\begin{eqnarray}
	L_{\nu + p} &=& a_p \, 	L_{\nu + p -1} + 	L_{\nu + p - 2}  \nonumber \\
	&=& \left(a_p a_{p-1} + 1 \right)  \, L_{\nu + p - 2} + a_p \, L_{\nu + p - 3}  \nonumber \\
	&=&   \left(a_p a_{p-1} a_{p-2} + a_{p-2} + a_p \right) \,  L_{\nu + p - 3} + \left(a_p a_{p-1} + 1 \right)  \, L_{\nu + p - 4} 
	= \cdots = \nonumber \\
	&=& v_p \, L_{\nu} + u_p \, L_{\nu-1}
	\label{eq057}
\end{eqnarray}
where $u_p/v_p = [0;a_1,\ldots,a_p]$ is the $p$--th convergent of $\alpha$. Dividing by $L_\nu$ we obtain finally
\begin{equation}
	\frac{L_{\nu+p}}{L_{\nu}} = u_p  	 \,   \frac{L_{\nu-1}}{L_{\nu}} + v_p 
	\label{equ058}
\end{equation}
Using again that we are considering $\nu = mp$ as multiple $p$ with $\nu \gg p$, then we can approximate $L_{\nu-1}/L_\nu = \tau_\nu \approx \beta$.
\begin{equation}
	\frac{L_{\nu-1}}{L_\nu} = \tau_\nu  = [0;a_p,\ldots,a_1,a_p,\ldots,a_1,\ldots,a_p,\ldots,a_1+\theta_A/\theta_B] \approx \beta
	\label{eq059}
\end{equation}
Plugging Eq. \eqref{eq059} into Eq.  \eqref{equ058}, the value of $\hat{	\tau}$ is finally
\begin{equation}
	\hat{	\tau} = \frac{1}{v_p + \beta \, u_p}
	\label{eq060}
\end{equation}
In order to prove that $\hat{	\sigma} \approx \hat{	\tau}$, the ratio $\hat{	\sigma} / \hat{	\tau}$ will be calculated using the derived forms above, \eqref{eq054} y \eqref{eq060}. 
\begin{equation}
	\frac{\hat{	\sigma}}{\hat{	\tau}} 
	= 	(-1)^p  \left(v_p - \frac{u_p}{\alpha}\right)  \, \left( v_p + \beta \, u_p  \right) 
	=  	(-1)^p \left[  v_p^2 - u_p^2 \, \frac{\beta}{\alpha} + u_pv_p \left(\beta -   \frac{1}{\alpha}\right) \right]
	\label{eq062}
\end{equation}  
This expression can be simplified even more making use of a known result concerning  periodic continued fractions. Indeed, it can be proved \cite{Olds-1963} that the following quadratic equation
\begin{equation}
	X^2 + \frac{v_p - u_{p-1}}{u_p} \, X - \frac{v_{p-1}}{u_p} = 0
	\label{eq063}
\end{equation} 
has $X_1=\beta$ y $X_2=-1/\alpha$ as roots. Thus, using the relationships between roots and polynomial coefficients, we have
\begin{equation}
	\frac{\beta}{\alpha} = \frac{v_{p-1}}{u_p} \quad , \quad \beta -   \frac{1}{\alpha} = - \frac{v_p - u_{p-1}}{u_p} 
	\label{eq064}
\end{equation}
Plugging this result into Eq.~\eqref{eq062} and after some algebra it yields
\begin{equation}
	\frac{\hat{	\sigma}}{\hat{	\tau}}  = 
	(-1)^p \left(  v_p \, u_{p-1} - u_p \, v_{p-1} \right) = (-1)^p \cdot (-1)^p = 1
	\label{eq065}
\end{equation}
where the identity of Eq.\eqref{eq005} has been invoked. Proved the fact that $\hat{	\sigma} = \hat{	\tau}$, we have finally from  Eq.~\eqref{eq052} that
\begin{equation}
	H(\hat{\sigma} \, k_{\nu+1}) = \hat{	\sigma}^2 \,   H(k_{\nu+1})
	\label{eq066}
\end{equation}
which demonstrates the quadratic decay of the Fourier intensities considering the full period of $p$ steps and thus $S(k) \sim k^4$. As above, the fact that the spectrum is formed by a singular set of Bragg peaks makes the cumulative intensities to behave under the same power-law, that is, enveloped as $Z(k) \sim k^4$.  Fig. \ref{fig04}(bottom) show the Fourier intensities $H(k)$ and their cumulative function $Z(k)$ for the system generated by the periodic continuous fraction $\alpha = [\overline{1,1,6,2}]$, with a period of 4 digits. According to theoretical derivations, the Bragg peaks associated with the sequence of dominant wavenumbers $\{k_\nu, \ 0 \leq \nu n\}$ are also arranged periodically on the logarithmic scale. The Fourier magnitudes are scaled under the same pattern that the density fluctuations every 4 steps, something that is clearly reflected in both plots. It is observed that the power-lay enveloping the $Z(k)$ function is exactly of order 4, validating the theoretical pattern derived in Eq. \eqref{eq066}.\\

In the two previous sections the cases of periodic irrational numbers have been considered. The general case of a tiling generated by any continued fraction is studied in the next section, showing that the global asymptotic exponent of the decreasing Fourier intensities towards $k\to 0$ is demonstrated to be quadratic. \\

\subsection{The general case: $\alpha = [0;a_1,\ldots,a_n]$}

After studying the specific cases seen in the previous two points, it is worth asking whether the detected behavior can be generalized to any quasiperiodic medium generated by a continued fraction $\alpha = [0;a_1,\ldots,a_n]$, exhibiting hyperuniform behavior and a structure factor that tends to zero according to a quartic law, i.e., $S(k) \sim k^4$ as $k \to 0$. It has been shown that, locally, differences in the values of the sequence $\{a_j\}$ are reflected in perturbations of the Fourier intensities, as observed in the numerical examples in Figs.~\ref{fig02} and \ref{fig04}. Thus, the exponent $1 + \log \tau_\nu / \log \sigma_\nu$, which affects the wavenumbers according to Eq.~\eqref{eq045}, may have high local values. However, the structure of the parameters $\tau_\nu$ and $\sigma_\nu$ themselves causes the slopes to be smoothed out somewhat in subsequent steps as the parameter $\nu$ progresses between $0 \leq \nu \leq n$. At this point, we will see that indeed the relationship between the Fourier coefficients at the first and last steps, i.e., $\nu=0$ and $\nu=n$, is approximately quadratic when $n \to \infty$. That is,
\begin{equation}
	\frac{H(k_n)}{H(k_0)} \approx \left(\frac{k_n}{k_0}\right)^\chi \quad , \quad n \to \infty
	\label{eq067}
\end{equation}
Without loss of generality, we will name again
\begin{equation}
	\hat{\sigma} = \sigma_{0} \, \sigma_{1} \, \cdots \, \sigma_{n-1} \ , \qquad
	\hat{\tau} = \tau_{0} \, \tau_{1} \, \cdots \, \tau_{n-1}
	\label{eq068}
\end{equation}
So that
\begin{eqnarray}
	k_{n} &=& \sigma_{n-1} \, k_{n-1} = \cdots = \sigma_{n-1} \cdots \sigma_1 \, \sigma_0 \, k_0 \quad , \quad \equiv \hat{\sigma} \, k_0 \nonumber \\
	H(k_n) &=& \sigma_{n-1} \, \tau_{n-1} \, H( k_{n-1}) \nonumber \\
	&=& \left( \sigma_{n-1} \, \cdots \, \sigma_{1} \, \sigma_{0} \right) \, \cdots \left( \tau_{n-1} \, \cdots \, \tau_{1} \, \tau_{0} \right) H(k_0) \nonumber \\
	&\equiv & \hat{\sigma} \, \hat{\tau} \, H(k_0)
	\label{eq069}
\end{eqnarray}
Using the expressions \eqref{eq034} and \eqref{eq047}, the values of $\hat{\sigma}$ and $\hat{\tau}$ can be simplified as
\begin{equation}
	\hat{	\sigma} = (-1)^n \, \frac{\xi_{n-1}}{\xi_{-1}} = \frac{1 }{v_n} \quad , \quad
	\hat{	\tau} = \frac{L_{-1}}{L_{n-1}} = \frac{A}{L_{n-1}} = \frac{A}{\tau_n \, L_{n}} = \frac{1}{ \tau_n (\alpha + \theta_A / \theta_B)} \frac{1}{v_n} 
	\label{eq070}
\end{equation}
Thus, we can then calculate the value of $\chi$ as
\begin{equation}
	\chi = \frac{\log H(k_n) - \log H(k_0)}{\log k_n - \log k_0 } = \frac{\log \hat{	\sigma} + \log \hat{	\tau} }{\log \hat{	\sigma}}= 2 + \frac{\log \left[ \tau_n (\alpha + \theta_A / \theta_B) \right] }{\log v_n} \approx  2 \qquad (n \to \infty)
\end{equation}
The above expression tends to 2 because the sequence $v_n$ of natural numbers increases indefinitely, while $\tau_n$ in general remains as less than unity. \\

\begin{figure}[H]%
	\begin{center}
		\includegraphics[width=12cm]{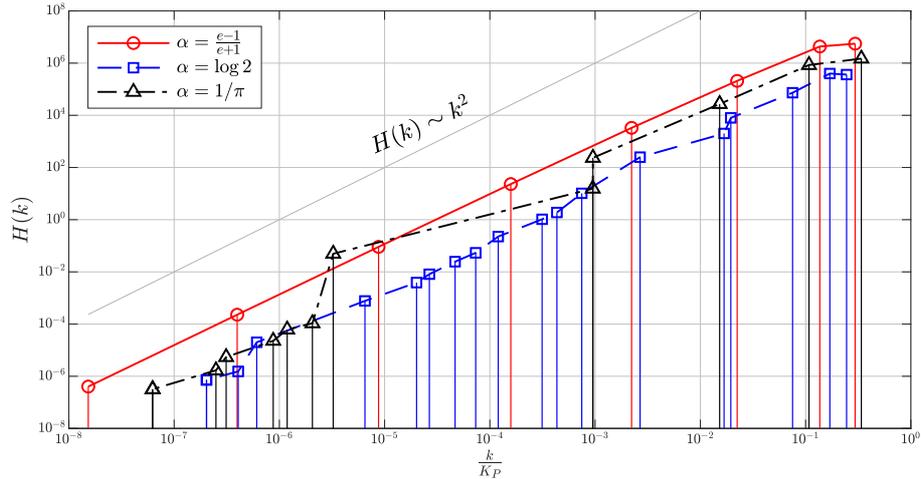} \\	
		\caption{Fourier intensities for three quasiperiodic tilings generated by $\alpha = \{\frac{e-1}{e+1},\log 2, 1/\pi \}$. The Fourier magnitudes been evaluated at the  dominant sequence of wavenumbers, which are no necesarly equal for the three tilings.}%
		\label{fig05}%
	\end{center}
\end{figure}

Several numerical examples have been carried out to verify this property, all of them showing a quadratic exponent in the trend toward the long-wavelength range. Three of them are illustrated in Fig. \ref{fig05}, generated with the following irrational numbers and their corresponding continued fractions
\begin{eqnarray}
	\alpha &=& \frac{e-1}{e+1} = [0;2,6,10,14,18,22,\ldots] = 0.4611715\ldots \nonumber \\
	\alpha &=& \ln 2 = [0;1,2,3,1,6,3,1,1,2,\ldots]= 0.69314718\ldots	 \nonumber \\
	\alpha &=& \frac{1}{\pi} =[0;3,7,15,1,292,1,1,1,2,1,3,1] = 0.31830988\ldots
	\label{eq077}
\end{eqnarray}
Fig. \ref{fig05} reveals the asymptotic behavior proved in the theoretical derivation. Although there may be significant fluctuations locally, as for example in the case of $1/\pi$, finally on average the hyperuniformity coefficient is common and equal to $\gamma = 3$.


\section{Conclusions}

In this paper the hyperuniformity of one-dimensional quasiperiodic lattices generated by continued fractions has been studied. Given any real number in the interval [0,1] as a continued fraction, we can construct a word or sequence from a binary alphabet, giving rise to quasiperiodic tilings. The studied media are constructed by word concatenation as one-dimensional quasiperiodical distributions of points. The Fourier intensities in the reciprocal space are recursively determined thus exploiting the quasiperiodic nature of the tiling. Among the entire spectrum of Bragg peaks, a sequence of wavenumbers, called {\em dominant sequence of wavenumbers}, has been identified, showing special properties related to the density fluctuations of the tiling. It has been proved that the pattern of decay of Fourier intensities at this sequence is quadratic regardless the continued fraction, meaning these media are strongly hyperuniform with exponent 3. The theoretical results have been validated and illustrated by means of several numerical examples.

\section*{Acknowledgments}

Ll. M. Garc\'ia-Raffi and M. L\'azaro are grateful for the partial support by the Grant PID2020-112759GB-I00 funded by MCIN/AEI/10.13039/501100011033. M. L\'azaro is grateful by the support of the Grant CIGE/2021/141 funded by Generalitat Valenciana (Emerging Research Groups).

\section*{References}
\bibliographystyle{elsarticle-num} 
\bibliography{bibliography}





\end{document}